\documentclass[a4paper,12pt]{article}
\usepackage{amsthm}
\usepackage{amssymb}
\usepackage{amsmath}
\usepackage{amsfonts}

\newtheorem{lemma}{Lemma}[section]
\newtheorem{theorem}{Theorem}[section]

\newtheorem{corollary}{Corollary}[section]

\def\b1{\mbox{\boldmath $1$}}

\newenvironment{demo*}{\vspace{3mm}\noindent{\bf Proof.}}{\hfill $\Box$ \vspace{3mm}}

\begin{document}

\baselineskip=20pt

\title{\bf  Alternative approach to the optimality of the threshold  strategy
for spectrally negative L\'evy processes}
\author{$^a$Ying Shen\footnote{
Corresponding author. Tel:+865374453221; fax:+865374455076
\newline
 {\it E-mail addresses:} ccyin@mail.qfnu.edu.cn (C. C. Yin), kcyuen@hku.hk (K. C.
 Yuen)},\ \ $^a$Chuancun Yin
 \ \ $^b$Kam Chuen Yuen,\\
{\normalsize\it $^a$School of Mathematical Sciences, Qufu Normal
University,}
{\normalsize\it Shandong 273165,\ China} \\
[3mm] {\normalsize\it $^b$Department of Statistics and Actuarial Science, The University of Hong Kong,} \\
{\normalsize\it Pokfulam Road, Hong Kong}\\
}
\date{}
\maketitle

 \vskip0.01cm
 \noindent{\large {\bf Abstract}}
 Consider the optimal dividend problem for an insurance company
 whose uncontrolled surplus precess evolves as a spectrally negative L\'evy process.
 We assume that dividends are paid to the shareholders according to admissible strategies whose
 dividend rate is bounded by a constant. The objective is to find a
 dividend policy so as to maximize the  expected discounted value of dividends which are paid to the
 shareholders until the company is ruined.   In this paper, we   shown that a threshold
strategy (also called refraction  strategy) forms an optimal strategy under the condition that the
L\'evy measure has a completely monotone density.

\noindent {2000 MR Subject Classification}: 60J51; 93E20; 91B30



\normalsize

\baselineskip=20pt

\section{Introduction}\label{intro}
The classical optimal dividend problem looks for the strategy that
maximizes the expected discounted dividend payments until ruin in an
insurance portfolio, which has recently received a lot of attention
in actuarial mathematics. This optimization problem was first
proposed by De Finetti [11] to reflect more realistically the
surplus cash flows in an insurance portfolio, who considered a
discrete time random walk with step sizes $\pm$ 1 and proved that
the optimal dividend strategy is a barrier strategy. Since then many
researchers have tried to address this optimality question under
more general and more realistic model assumptions and until nowadays
this turns out to be a rich and challenging field of research that
needs the combination of tools from analysis, probability and
stochastic control. For the classical compound Poisson risk model,
this problem was solved by Gerber in [15] via a limit of an
associated discrete problem. Recently, this optimal dividend problem
in the classical compound Poisson risk model and also included a
general reinsurance strategy as a second control possibility was
taken up again by  Azcue and Muler [7], who used stochastic control
theory and viscosity solutions.   For all these cases in general a
band strategy turns out to be optimal among all admissible
strategies.  In particular, for exponentially distributed claim
sizes this optimal strategy simplifies to a barrier strategy. In
Albrecher and Thonhauser [1] it is shown that the optimality of
barrier strategies in the classical model with exponential claims
still holds if there is a constant force of interest.  Avram et al.
[5] considered the case where the risk process is given by a general
spectrally negative L\'evy process and gave a sufficient condition
involving the generator of the L\'evy process for optimality of the
barrier strategy. Recently, Loeffen [29] showed that barrier
strategy is optimal among all admissible strategies for general
spectrally negative L\'evy risk processes with completely monotone
jump density, and Kyprianou et al. [26] relaxed this condition on
the jump density to log-convexity. More recent paper Azcue and Muler
[8] examines the analogous questions in the compound Poisson risk
model with investment. The corresponding problem in the case of a
diffusion risk process was completely solved by Shreve et al. [32]
and a barrier strategy was identified to be optimal. The special
case of constant drift and diffusion coefficient was then solved
again by slightly different means in Jeanblanc-Picqu\'e and Shiryaev
[21] and Asmussen and Taksar [4].

Band strategy (and barrier strategy) often serve as candidates for
the optimal strategy when the dividend rate is unrestricted.
However, the resulting dividend stream is far from practical
application. In many circumstances this is not desirable.
Furthermore, if a band strategy is applied, ultimate ruin of the
company is certain. Motivated by this fact, other dividend
strategies such as threshold strategies, linear and nonlinear
barrier strategies and multi-layer  strategies have been studied.
 Asmussen and Taksar [4] postulated a bounded
dividend rate and showed that the optimal dividend strategy is a
threshold strategy in Brownian motion risk models, that is,
dividends should be paid out at the maximal admissible rate as soon
as the surplus exceeds a certain threshold. Some calculations for
this model can be found in [16]. In the compound Poisson risk model,
Gerber and Shiu [17] showed that the optimal dividend strategy is a
bang bang strategy.  In particular, for exponentially distributed
claim sizes this optimal strategy simplifies to a threshold
strategy. Motivated by  Gerber and Shiu [17], Fang and Wu [13] study
the analogous questions in the compound Poisson risk model with
constant interest, for the case of an exponential claim amount
distribution, it is shown that the optimal dividend strategy is a
threshold strategy. More recently, Fang and Wu [14] examine the same
problem for the Brownian motion risk model with interest. In a very
recent paper,   Kyprianou, Loeffen and P\'erez [28] have shown that
a refraction  strategy (also called threshold strategy) forms an
optimal strategy under the condition that the L\'evy measure has a
completely monotone density. See Albrecher and Thonhauser [2],
 Avanzi [6] and  Schmidli [31] for nice
surveys on this subject. The purpose of this paper is to re-examine
the analogous questions in a general spectrally negative L\'evy
process risk model.

The rest of the paper is organized as follows.  In Section 2, we
state the problem and recall some preliminaries on spectrally
negative L\'evy processes.  In Section 3, we will show that the
optimal value function of the dividends can be characterized by the
Hamilton-Jacobi-Bellman (HJB) equation and give a verification
result for optimality.   In Section 4 we discuss the threshold
strategies. Explicit expressions and the integro-differential
equations for the   expected discounted value of dividend payments
are obtained, and in Section 5 we present the main results.

  \vskip 0.2cm
\section{  Problem setting}\label{problem}
\setcounter{equation}{0} Suppose that $X =(X(t) : t\ge 0)$ is a
spectrally negative L\'evy process with probabilities $\{P_x : x\in
\Bbb{R}\}$ such that $X(0)=x$ with probability one, where we write
$P=P_0$. Let $E_x$ be the expectation with respect to $P_x$ and
write $E=E_0$.  Let $\{{\cal{F}}_t : t\ge 0\}$ be the natural
filtration satisfying the usual assumptions. Since the jumps of a
spectrally negative L\'evy process are all non-positive, for
convenience, we choose the L\'evy measure to have mass only on the
positive instead of the negative half line. The Laplace exponent of
$X$ is given by
\begin{equation}
\psi (\theta) = a\theta + \frac1 2\sigma^2\theta^2
-\int_0^{\infty}(1-e^{-\theta x}-\theta x\text{\bf
1}_{\{0<x<1\}})\Pi (dx) \label{problem-eq1}
\end{equation}
 where $a\in\Bbb{R}$, $\sigma\ge
0$ and $\Pi$ is a measure on $(0,\infty)$ satisfying
$\int_0^{\infty}(1\wedge x^2)\Pi(dx)<\infty$ and is called the
L\'evy measure. The characteristics $(a, \sigma^2, \Pi)$ are called
the L\'evy triplet of the process and completely determines its law.
$\psi$ is strictly convex on $(0,\infty)$ and satisfies $\psi(0+) =
0$, $\psi(\infty) = \infty$ and $\psi'(0+)=E X(1)$.  If $\sigma^2>0$
and $\Pi=0$, then the process is a Brownian motion; When
$\sigma^2=0$ and $\int_0^{\infty}\Pi(dx)<\infty$, the process is a
compound Poisson process; When $\sigma^2=0$,
$\int_0^{\infty}\Pi(dx)=\infty$ and $\int_0^{\infty}(1\wedge
x)\Pi(dx)<\infty$,    the process has an infinite number of small
jumps but is of finite variation; When $\sigma^2=0$,
$\int_0^{\infty}\Pi(dx)=\infty$ and $\int_0^{\infty}(1\wedge
x)\Pi(dx)=\infty$,    the process has infinitely many   jumps and is
of  unbounded variation. In a word, such a L\'evy process has
bounded variation if and only if $\sigma=0$ and
$\int^{1}_{0}x\Pi(\text{d}x)<\infty$. In this case the L\'evy
exponent can  be re-expressed as
 $$\psi(\beta)=c\beta-\int_0^{\infty}(1-\text{e}^{-\beta
 x})\Pi(\text{d}x),$$
 where $c=a+\int^{1}_0 x\Pi(\text{d}x)$ is known as the drift
 coefficient.
  If $\sigma^2>0, X$ is said to have a Gaussian component.

 In this paper, we shall only consider the case that $\Pi$ is absolutely continuous with
 respect to Lebesgue measure, in which case we shall refer to its density as $\pi$.

We recall from Kyprianou [24] that for each $q\ge 0$ there exits a
continuous and increasing function $W^{(q)}:\Bbb{R}\rightarrow
[0,\infty)$, called the $q$-scale function defined in such a way
that $W^{(q)}(x) = 0$ for all $x < 0$ and on $[0,\infty)$ its
Laplace transform is given by
\begin{equation}
\int_0^{\infty}\text{e}^{-\theta
x}W^{(q)}(x)dx=\frac{1}{\psi(\theta)-q},\; \theta
>\Phi(q), \label{problem-eq2}
\end{equation}
where $\Phi(q)=\sup\{\theta\ge 0: \psi(\theta)=q\}$ is the
right-inverse of $\psi$. We shall write $W$ in place of $W^{(0)}$
and call this the scale function rather than the 0-scale function.

The q-scale function takes its name from the identity
$$E_x(e^{-q\tau_a^+}{\text {\bf
 1}}(\tau_a^+<\tau_0^-)=\frac{W^{(q)}(x)}{W^{(q)}(a)},$$ where
 $$\tau^+_a=\inf\{t\ge 0: X(t)>a \},\;\;\;\tau_0^-=\inf\{t\ge 0: X(t)<0 \}.$$

  The following facts about   the scale functions are taken from [10, 26]. If X has paths of bounded
variation then, for all $q\ge 0$, $W^{(q)}|_{(0,\infty)}\in
C^1(0,\infty)$ if and only if $\Pi$ has no atoms. In the case that
$X$ has paths of unbounded variation, it is known that, for all
$q\ge 0$, $W^{(q)}|_{(0,\infty)}\in C^1(0,\infty)$. Moreover if
$\sigma> 0$ then $C^1(0,\infty)$ may be replaced by $C^2(0,\infty)$.
Further, if the L\'evy measure has a density, then the scale
functions are always differentiable. In particular,   if $\pi$ is
completely monotone then $W^{(q)}|_{(0,\infty)}\in
C^{\infty}(0,\infty)$. It is well known that $W^{(\delta)}(0+)=1/c$
when $X$ has paths of bounded variation. Otherwise
$W^{(\delta)}(0+)=0$ for the case of unbounded variation. In all
cases, if $EX(1)>0$, then   $W(\infty)=1/EX(1)$. If $q>0$, then
$W^{(q)}(x)\sim e^{\Phi(q)x}/\psi'(\Phi(q))$ as $x\to\infty$.

Spectrally negative L\'evy processes have been considered recently
in [12, 18,  20, 23, 25,  34],  among others, in the context of
insurance risk models. It is assumed that, in the absence of
dividends, the surplus of a company at time $t$ is $X(t)$. We assume
now that the company pay dividends to its shareholders according to
some strategy. Let $\xi=\{L_t^{\xi}:t\ge 0\}$ be a dividend strategy
consisting of a left-continuous non-negative non-decreasing process
adapted to the filtration $\{{\cal{F}}_t:t\ge 0\}$ of $X$.
$L_t^{\xi}$ represents the cumulative dividends paid out up to time
$t$ under the control $\xi$ by the insurance company whose risk
process is modelled by $X$. We define the controlled risk process
$U^{\xi}=\{U^{\xi}(t):t\ge 0\}$ by $U^{\xi}(t)=X(t)-L_t^{\xi}$. Let
$T=\inf\{t>0: U^{\xi}(t)<0\}$ be the ruin time,  and define the
value function of a dividend strategy $\xi$ by
$$V_{\xi}(x)=E_x\left(\int_0^T\text{e}^{-q
t}dL_t^{\xi}\right),$$ where $q>0$ is the discounted rate.

A dividend strategy is called admissible if
$L_{t+}^{\xi}-L_t^{\xi}\le U^{\xi}(t)\vee 0$ for $t<T$, in other
words the lump sum dividend payment is smaller than the size of the
available capitals.  Let $\Xi$ be the set of all admissible dividend
policies. The control problem consists of   solving the following
stochastic control problem:
$$V_*(x)=\sup_{\xi\in\Xi}V_{\xi}(x),$$
and, if it exists,  to find a strategy $\xi^*\in\Xi$ such that
$V_{\xi^*}(x)=V_*(x)$ for all $x\ge 0$.

In this paper, we assume that the admissible dividend rate is $r(t)$
at time $t$ which is bounded by a constant $\alpha$. In the sequels,
we assume that $0<\alpha<a+\int_0^1 x\pi(x)dx$ if $X$ has paths of
bounded variation. Under this additional constraint, we will show
that  if the  L\'evy measure $\Pi$ has a completely monotone density
and that $\delta>0$, then the optimal dividend strategy is formed by
a threshold strategy.

\setcounter{equation}{0}
\section{ The HJB equation and verification of optimality }\label{HJB}
\setcounter{equation}{0}

 Let $$V(x)=\sup_{r(\cdot)}V_r(x),$$ where the supremum is taken over all control process $r(t)$
 which are  admissible according to the constraints. $V_r$ is the value function
 when the admissible dividend rate is $r(t)$ at time $t$ and
   $V$ is called the optimal value function. Suppose $V$ is twice
   continuously differential on $(0,\infty)$ when the process $X$ is
   of unbounded variation and is  continuously differential on $(0,\infty)$ when the process $X$ is
   of bounded variation.
Standard Markovian arguments yield that $V$ satisfies the following
Hamilton-Jacobi-Bellman (HJB) equation (the proof is similar to the
one in Azcue and Muler [8]):
\begin{equation}
\max_{0\le r\le \alpha} [1-V'(x)]r+\Gamma V(x,b)-\delta V(x,b)=0,\;
x\ge 0, \label{HJB-eq1 }
\end{equation}
where
\begin{equation}
\begin{array}{lll}
 \Gamma V(x,b)&=&\frac{1} {2}\sigma^2 V''(x,b)
  +a V'(x,b)\\
  &&+\int_0^{\infty}[V(x-y,b)-V(x,b)+V'(x,b)y\text{\bf
1}_{(0<y<1)}]\pi(y)dy. \label{HJB-eq2 }
\end{array}
 \end{equation}
 From above expression, as in Gerber and Shiu [17],  we see
that at time $t\in (0,T)$, the optimal dividend rate is
$$r=0  \;\; {\text{\rm if}}\;\; V'(U^{\xi}(t-))>1,$$
$$r=\alpha   \;\; {\text{\rm if}}\;\;  V'(U^{\xi}(t-))<1.$$ If $V'(U^{\xi}(t-))=1$, the dividend rate
$r$ can be any value between $0$ and $\alpha$. In particular, the
optimal dividend rate at time $0$ is
 $$r=0 \;\; {\text{\rm if}}\;\; V'(x)>1,$$
$$r=\alpha   \;\; {\text{\rm if}}\;\; V'(x)<1.$$
Thus, the company should either pay nothing, or the maximum
possible. This called a bang bang strategy.

 Now we show that a
strategy $\xi$ with $\nu(x)\equiv V_{\xi}(x)$ is smooth enough and
satisfying the HJB equation  (\ref{HJB-eq1 }) is indeed an optimal
strategy.

Consider any other dividend strategy, with dividend rate $r(t)$ and
surplus $\tilde{X}(t)$ at time $t$. We claim that
\begin{equation}
E\left(\int_0^T e^{-\delta
t}r(t)dt|\tilde{X}(0)=x\right)\le\nu(x).\label{HJB-eq3}
\end{equation}

From this, it follows that $\nu(x)=V(x)$, and hence the given
strategy $\xi$ is optimal.

To prove inequality  (\ref{HJB-eq3}), we consider the martingale
$$e^{-\delta t}\nu(\tilde{X}(t))-\int_0^t e^{-\delta
s}\left[(\Gamma-\delta) \nu(\tilde{X}(s))-r(s)
\nu'(\tilde{X}(s))\right]ds,$$ which can be shown by It\^{o}'s
formula for semimartingale. From optimal sampling theorem, we have
$$E\left(e^{-\delta (t\wedge T)}\nu(\tilde{X}(t\wedge T))-\int_0^{t\wedge T} e^{-\delta
s}\left[(\Gamma-\delta) \nu(\tilde{X}(s))-r(s)
\nu'(\tilde{X}(s))\right]ds|\tilde{X}(0)=x\right)=\nu (x),$$ which
implies
$$-E\left(\int_0^{t\wedge T} e^{-\delta
s}\left[(\Gamma-\delta) \nu(\tilde{X}(s))-r(s)
\nu'(\tilde{X}(s))\right]ds|\tilde{X}(0)=x\right)\le \nu (x),$$ since
$$E\left(e^{-\delta (t\wedge T)}\nu(\tilde{X}(t\wedge T))|\tilde{X}(0)=x\right)\ge
0.$$ Because the function $\nu(x)$ satisfies the HJB equation
(\ref{HJB-eq1 }), we have
$$r(s)+(\Gamma-\delta) \nu(\tilde{X}(s))-r(s) \nu'(\tilde{X}(s))\le
0.$$ Thus
 \begin{eqnarray*}
  &E&\left(\int_0^{t\wedge T} e^{-\delta
s}r(s)ds|\tilde{X}(0)=x\right)\\
&&\le -E\left(\int_0^{t\wedge T} e^{-\delta s}\left[ (\Gamma-\delta)
\nu(\tilde{X}(s))-r(s) \nu'(\tilde{X}(s))\right]
ds|\tilde{X}(0)=x\right)\le \nu(x). \end{eqnarray*}
 Letting
$t\to\infty$ yields  (\ref{HJB-eq3}).

\setcounter{equation}{0}
\section{ Threshold dividend strategies}\label{Thre}

In this section, we study  the threshold strategy. We assume that
the company pays dividends according to the following strategy
governed by parameters $b>0$ and $\alpha>0$. Whenever the modified
surplus is below the threshold level $b$, no dividends are paid.
However,   when the surplus is above this threshold level, dividends
are paid continuously at a constant rate $\alpha$  that does not
exceed the premium rate $c$. Note that   if $\alpha=c$, we have a
barrier strategy again. We define the modified   risk process
$U_b=\{U_b(t):t\ge 0\}$ by $U_b(t)=X(t)-D_b(t)$, where
$D_b(t)=\alpha\int_0^t {\text{\bf 1}} (U_b(t)>b)dt$. The existence
of such a process  and some conclusions on fluctuation identities
can be found in  Kyprianou and Loeffen [27].  Let $D_b$ denote the
present value of all dividends until time of ruin $T$,
$$D_b=\alpha\int_0^T e^{-\delta t} {\text{\bf 1}} (U_b(t)>b)dt$$  where
   $T=\inf\{t>0: U_b(t)<0\}$
with $T=\infty$ if $U_b(t)\ge 0$ for all $t\ge 0$. Here  $\delta>0$
is the discount factor. Denote by   $V(x,b)$ the expected discounted
value of dividend payments, that is,
$$V(x,b)=E(D_b|U_b(0)=x).$$
Clearly, $0\le V(x,b)\le \frac{\alpha}{\delta}$ and
$\lim_{x\to\infty}V(x,b)=\frac{\alpha}{\delta}$.

 Define the first passage times, with the convention $\inf
\emptyset=\infty$,
$$T^+_b=\inf\{t\ge 0: U_b(t)>b \},\;\;\;T_b^-=\inf\{t\ge 0: U_b(t)\le b \}.$$

Let $Y=\{Y(t):=X(t)-\alpha t\}_{t\ge 0}$. For each $\delta\ge 0$,
$W^{(\delta)}$ and $Z^{(\delta)}$ are the $\delta$-scale functions
associated with $X$ and $W_{*}^{(\delta)}$ and $Z_{*}^{(\delta)}$
are the $\delta$-scale functions associated with $Y$. Further,
$\Psi$ is defined as the right inverse of the Laplace exponent of
$Y$ so that
$$\Psi(\delta)=\sup\{\theta\ge 0: \psi (\theta)-\alpha \theta=\delta\}.$$
  \begin{theorem}\label{thrm4-1}   Assume $W^{(\delta)}$ is continuously
differentiable on $(0,\infty)$.

 (1)\; For $0\le x\le b$, we have
\begin{equation}
V(x,b)=V(b,b)\frac{W^{(\delta)}(x)}{W^{(\delta)}(b)}.\label{Thre-eq1}
\end{equation}
(2)\; For $x>b$, we have
 \begin{eqnarray}
  V(x,b)&=&-\alpha\int_0^{x-b}W_{*}^{(\delta)}(y)dy
 +\frac{\alpha}{\Psi(\delta)}W_{*}^{(\delta)}(x-b)\nonumber\\
&&+\frac{\sigma ^2}{2}V(b,b)\left(W_{*}{^{(\delta)}}'(x-b)-\Psi(\delta)W_{*}^{(\delta)}(x-b)\right)\nonumber\\
&&+\frac{V(b,b)}{W^{(\delta)}(b)}\int_b^{\infty}dy\int^b_{-\infty}
u_{*}^{(\delta)}(x-b,y-b)\pi(y-z)W^{(\delta)}(z)dz.
\label{Thre-eq2}
 \end{eqnarray}
 where
$$u_{*}^{(\delta)}(x,y)=W_{*}^{(\delta)}(x)e^{-\Psi(\delta)y}-W_{*}^{(\delta)}(x-y).$$
\end{theorem}

{\bf Proof} (1). For $0\le x\le b$, using the strong Markov property
of $U$ at $T_b^+$, we have
$$V(x,b)=V(b,b)E_x(e^{-\delta T_b^+}{\text{\bf1}}(T_b^+<T)),$$ and (\ref{Thre-eq1}) follows
since
$$E_x(e^{-\delta T_b^+}{\text{\bf1}}(T_b^+<T))=\frac{W^{(\delta)}(x)}{W^{(\delta)}(b)}.$$

(2). For $x>b$,  using the strong Markov property of $U_b$ at
$T_b^-$, we have
\begin{eqnarray}
 V(x,b)&=&\frac{\alpha}{\delta}P_x(T_b^-=\infty)
+\alpha E_x\left(\int_0^{T_b^-} e^{-\delta u}du,
T_b^-<\infty\right) \nonumber\\
&&+\alpha E_x\left(\int_{T_b^-}^{T} {\text{\bf1}}(U_b(u)>
b)e^{-\delta
u}du, T_b^-<\infty\right)\nonumber\\
&=&\frac{\alpha}{\delta}P_x(T_b^-=\infty)+\frac{\alpha}{\delta}
 E_x \left((1-e^{-\delta T_b^-}),T_b^-<\infty\right)\nonumber\\
&&+E_x \left( e^{-\delta T_b^-}V(U_b(T_b^-),b),
T_b^-<\infty\right)\nonumber\\
&=&\frac{\alpha}{\delta}-\frac{\alpha}{\delta}
 E_x \left(e^{-\delta T_b^-},T_b^-<\infty\right)\nonumber\\
&&+ E_x \left( e^{-\delta T_b^-}V(U_b(T_b^-),b)
,T_b^-<\infty\right). \label{Thre-eq3}
 \end{eqnarray}
 Note that
\begin{eqnarray*}
 E_x \left( e^{-\delta
T_b^-}V(U_b(T_b^-),b), T_b^-<\infty\right)&=& V(b,b)E_x \left(
e^{-\delta
T_b^-}, U_b(T_b^-)=b)\right)\\
&&+\int_b^{\infty}dy\int^b_{-\infty}
u_{*}^{(\delta)}(x-b,y-b)\pi(y-z)V(z,b)dz\\
&=&\frac{\sigma ^2}{2}V(b,b)\left(W_{*}{^{(\delta)}}'(x-b)-\Psi(\delta)W_{*}^{(\delta)}(x-b)\right)\\
&&+\int_b^{\infty}dy\int^b_{-\infty}
u_{*}^{(\delta)}(x-b,y-b)\pi(y-z)V(z,b)dz. \end{eqnarray*}

In particular,
$$
  E_x \left( e^{-\delta T_b^-}, T_b^-<\infty\right)
 = Z_{*}^{(\delta)}(x-b)-\frac{\delta}{\Psi(\delta)}W_{*}^{(\delta)}(x-b),
 $$
where
$$Z_{*}^{(\delta)}(x)=1+\delta\int_0^x W_{*}^{(\delta)}(y)dy.$$
 It
follows that
 \begin{eqnarray*} V(x,b)&=&-\alpha\int_0^{x-b}W_{*}^{(\delta)}(y)dy
 +\frac{\alpha}{\Psi(\delta)}W_{*}^{(\delta)}(x-b)\\
&&+\frac{\sigma ^2}{2}V(b,b)\left(W_{*}{^{(\delta)}}'(x-b)-\Psi(\delta)W_{*}^{(\delta)}(x-b)\right)\\
&&+\int_b^{\infty}dy\int^b_{-\infty}
u_{*}^{(\delta)}(x-b,y-b)\pi(y-z)V(z,b)dz.
\end{eqnarray*}
 Putting  (\ref{Thre-eq1}) into above expression leads to  (\ref{Thre-eq2}).

The following result agrees with the result of Kyprianou and Loeffen
[27, (10.25)].
\begin{corollary}\label{coro4-1} Suppose  $X$ has paths of bounded
variation and let $0<\alpha<c$, where $c=a+\int_0^1 x\Pi(dx)$.

 (1)\; For $0\le x\le b$, we have
\begin{equation}
V(x,b)=\frac{W^{(\delta)}(x)}{\Psi(\delta)e^{\Psi(\delta)b}\int_b^{\infty}e^{-\Psi(\delta)z}
{W^{(\delta)}}'(z)dz}.\label{Thre-eq4}
\end{equation}
(2)\; For $x>b$, we have
\begin{equation}
V(x,b)=-\alpha\int_0^{x-b}W_{*}^{(\delta)}(y)dy+\frac{W^{(\delta)}+\alpha\int_b^x
W_{*}^{(\delta)}(x-y){W^{(\delta)}}'(y)dy}{\Psi(\delta)e^{\Psi(\delta)b}\int_b^{\infty}e^{-\Psi(\delta)z}
{W^{(\delta)}}'(z)dz}. \label{Thre-eq5}
\end{equation}
\end{corollary}
{\bf Proof} (1).  It follows from Kyprianou and Loeffen [27, (4.10)]
that, for $x>b$,
\begin{eqnarray*}
&&\int_b^{\infty}dy\int^b_{-\infty}
u_{*}^{(\delta)}(x-b,y-b)\pi(y-z)W^{(\delta)}(z)dz\\
&&= W^{(\delta)}(x)+\alpha\int_b^x
W_{*}^{(\delta)}(x-z){W^{(\delta)}}'(z)dz\\
&&-\alpha
W_{*}^{(\delta)}(x-b)e^{\Psi(\delta)b}\int_b^{\infty}e^{-\Psi(\delta)z}
{W^{(\delta)}}'(z)dz.
\end{eqnarray*}
Substituting this into   (\ref{Thre-eq2}) and letting $x\to b$ we
find that
$$V(b,b)=\frac{W^{(\delta)}(b)}{\Psi(\delta)e^{\Psi(\delta)b}\int_b^{\infty}e^{-\Psi(\delta)z}
{W^{(\delta)}}'(z)dz}.$$ This, together with  (\ref{Thre-eq1}) and
(\ref{Thre-eq2}), leads to (\ref{Thre-eq4}) and (\ref{Thre-eq5}).

The next result was obtained by Gerber and Shiu [17] for the
compound Poisson model:

  \begin{theorem}\label{thrm4-2} Suppose that $X$ has no Gaussian component.
Then,  as a function of $x$, $V(x,b)$ satisfies the following
integro-differential equations:
 \begin{equation}
\begin{array}{lll}
 a V'(x,b)
&+&\int_0^{\infty}[V(x-y,b)-V(x,b)+V'(x,b)y\text{\bf
1}_{(0<y<1)}]\pi(y)dy\\
&=& \delta V(x,b), \; 0<x<b, \label{Thre-eq6} \end{array}
\end{equation}

\begin{equation}
\begin{array}{lll}
 (a-\alpha) V'(x,b)
&+&\int_0^{\infty}[V(x-y,b)-V(x,b)+V'(x,b)y\text{\bf
1}_{(0<y<1)}]\pi(y)dy\\
&=& \delta V(x,b)-\alpha, \; x>b, \label{Thre-eq7} \end{array}
\end{equation}
 with the continuity condition
$V(b-,b)=V(b+,b)=V(b,b)$. Moreover, if $X$ has paths of bounded
variation then
$$cV'(b-,b)=(c-\alpha)V'(b+,b)+\alpha;$$ If $X$ has paths of
unbounded variation then
$$V'(b-,b)=V'(b+,b),$$ where $c=a+\int_0^1
x\pi(x)dx$.
\end{theorem}
{\bf Proof} Equations (\ref{Thre-eq6}) and (\ref{Thre-eq7}) can be
proved by Ito's formula. It follows from  (\ref{Thre-eq1}) that
$V(b-,b)=V(b,b)$, since $W^{(\delta)}$ is continuous at $b$.   From
 (\ref{Thre-eq2}) we see that $V(x,b)$ is differential on $[b,\infty)$, where we
at $x=b$ mean the right-hand derivative. Consequently,
$V(b+,b)=V(b,b)$. This proves the continuity of $V$ at $b$. If $X$
has paths of bounded variation, then the derivative, $V'(x,b)$, is
not necessarily continuous at $x=b$. In fact, it follows from
equations  (\ref{Thre-eq6}) and (\ref{Thre-eq7}) that
$$cV'(b-,b)=(c-\alpha)V'(b+,b)+\alpha.$$
 If $X$ has paths of unbounded variation,  let $\Pi_n$ be measures on $(1/n,\infty)$:
 $$\Pi_n(dx)=\Pi(dx)\text{\bf
1}_{(1/n,\infty)},\;\;n\ge 1.$$
 Then $\int_0^1 x\Pi_n(dx)<\infty$. This is to say that a process
 $X_n$ with the L\'evy measure $\Pi_n$ has paths of bounded
 variation. From above we get
$$c_n V_n '(b-,b)=(c_n-\alpha)V_n '(b+,b)+\alpha,$$
where $c_n=a+\int_0^1 x\pi_n(x)dx$. Letting $n\to\infty$ yields
$$V'(b-,b)=V'(b+,b).$$ This ends the proof of Theorem 4.2.

The next result was obtained by Wan [33] for the compound Poisson
model perturbed by diffusion:

  \begin{theorem}\label{thrm4-3}  Suppose that $X$ has
a Gaussian component $\sigma>0$. Then,  as a function of $x$,
$V(x,b)$ satisfies the following integro-differential equations:
 \begin{equation}
\begin{array}{lll}\frac{1} {2}\sigma^2 V''(x,b)&+&a V'(x,b)
+\int_0^{\infty}[V(x-y,b)-V(x,b)+V'(x,b)y\text{\bf
1}_{(0<y<1)}]\pi(y)dy\\
&=& \delta V(x,b), \; 0<x<b,
\label{Thre-eq8}\end{array}\end{equation}
  \begin{equation}
\begin{array}{lll}  \frac{1} {2}\sigma^2 V''(x,b)&+&(a-\alpha) V'(x,b)
+\int_0^{\infty}[V(x-y,b)-V(x,b)+V'(x,b)y\text{\bf
1}_{(0<y<1)}]\pi(y)dy\\
&&= \delta V(x,b)-\alpha, \; x>b, \label{Thre-eq9}
\end{array}\end{equation}

with the boundary conditions $V(0,b)=0$. Moreover
$$V(b-,b)=V(b+,b)=V(b,b),\; V'(b-,b)=V'(b+,b).$$
\end{theorem}
{\bf Proof} Equations  (\ref{Thre-eq8}) and (\ref{Thre-eq9})  can be
proved by Ito's formula. If $U_b(0)=0$, because $\sigma>0$, ruin is
immediate and no dividend is paid, so we have $V(0,b)=0$. It follows
from  (\ref{Thre-eq1}) that $V(b-,b)=V(b,b)$, since $W^{(\delta)}$
is continuous at $b$. From  (\ref{Thre-eq3}) we have
$$V(b+,b)\le\frac{\alpha}{\delta} \left(1-E_b (e^{-\delta T_b^-})\right)+
V(b,b)E_b \left( e^{-\delta T_b^-}\right)=V(b,b),$$ where we have used the fact $P_b(T_b^-=0)=1$. Thus
$V(b+,b)=V(b,b)$. This proves the continuity of $V$ at $b$.

In the following, we prove that $\{\sigma B(t);t\geq0\}$ can  be
approximated by the process $\{c_{\varepsilon}t-\varepsilon
N_{\varepsilon}(t);\ t\geq0\}$, where $N_{\varepsilon}(t)$ is a
Poisson process with parameter $\lambda_{\varepsilon}>0,$ and
$c_{\varepsilon}>0$ is a constant. Now, we choose $\varepsilon,\
\lambda_{\varepsilon},$ and $c_{\varepsilon}$ such that ${Var}
[\varepsilon N_{\varepsilon}(t)]=\sigma^{2}t$ and
$E[c_{\varepsilon}t-\varepsilon N_{\varepsilon}(t)]=0.$ These two
conditions yield $\lambda_{\varepsilon}=\sigma^{2}/\varepsilon^{2}$
and $c_{\varepsilon}=\sigma^{2}/\varepsilon.$ It is easy to prove
that, when $\varepsilon\rightarrow0^{+},\
E[e^{z(c_{\varepsilon}t-\varepsilon N_{\varepsilon}(t))}]\rightarrow
e^{z^{2}\sigma^{2}t/2}.$ This shows that the process
$\{c_{\varepsilon}t-\varepsilon N_{\varepsilon}(t);\ t\geq0\}$
converges weakly to the  process $\{\sigma B(t);t\geq0\}$. It
follows that the L\'evy process $X$ with L\'evy triplet $(a, \sigma,
\Pi)$ can be approximated by the L\'evy process $X_{\varepsilon}$
with L\'evy triplet $(a+c_{\varepsilon}, 0, \Pi_{\varepsilon})$,
where $\Pi_{\varepsilon}=\Pi+\text{\bf 1}_{(x\ge \varepsilon)}$.
Therefore, by Theorem 4.2, for example, in the case of bounded
variation, we have
$$(c+c_{\varepsilon})V_{\varepsilon}'(b-,b)=(c+c_{\varepsilon}-\alpha)V_{\varepsilon}'(b+,b)+\alpha.$$
Letting $\varepsilon\to 0$ and noting that $c,\alpha,
V_{\varepsilon}'(b-,b)$ and $V_{\varepsilon}(b+,b)$ are bounded, and
$\lim_{\varepsilon\to 0}V_{\varepsilon}=V$, yields
$$V'(b-,b)=V'(b+,b).$$ This ends the proof of Theorem 4.3.

\setcounter{equation}{0}
\section{Optimal dividend strategies}\label{opt}

In some situations the optimal dividend strategy is a threshold
strategy. It is easy to  see that if $V'(x,0)<1$ for $x>0$, then the
threshold strategy with $b^*=0$ is optimal, if $V'(x,b^*)>1$ for
$x<b^*$ and $V'(x,b^*)<1$ for $x>b^*$, then the threshold strategy
with $b^*>0$ is optimal. The optimal threshold $b^*$ can be obtained by
$V'(b^*,b^*)=1$. In fact,  it is obvious for the case $\sigma>0$,
since $V'(x,b)$ is a continuous function of $x$ on $(0,\infty)$;
For the case $\sigma=0$, using the the same argument as in  Gerber
and Shiu [17], the result follows.  From those facts one sees that
if $V(x,b^*)$ is a continuously differentiable concave function on
$(0, \infty)$, then the optimal dividend strategy is a threshold
strategy.

 We now review  definitions and some properties of
logconvex functions and  completely monotone functions. We  refer
the readers  to [3, 9] for more details.

A function $f$ defined on an convex subset of a real vector space
and taking positive values is said to be logarithmically convex if
$\log(f(x))$ is a convex function of $x$. It is easy to see that a
logarithmically convex function is a convex function, but the
converse is not always true. For example $f(x)=x^2$ is a convex
function, but $\log(f(x))=2\log |x| $ is not a convex function and
thus $f(x) = x^2$ is not logarithmically convex.

Recall that a $f\in C^{\infty}(0,\infty)$ with $f\ge 0$ is
completely monotone if its derivatives alternate in sign, i.e.
$(-1)^n f^{(n)}\ge 0$ for all $n\in \Bbb{N}$.

Note that the class of logconvex functions contains the class of
completely monotone functions. In fact, any completely monotone
function  is both nonincreasing and logconvex.

 Some distributions with  completely monotone
density functions are (see  [9, 29]):

$\bullet$\;Weibull distribution with density: $f(x)=cr
x^{r-1}\text{e}^{-c x^r}, \;x>0,$ with $c>0$ and $0<r<1$.

 $\bullet$\;Pareto distribution with density: $f(x)=\alpha (1+x)^{-\alpha-1},
\;x>0,$ with $\alpha>0$.

 $\bullet$\;Mixture of exponential densities: $f(x)=\sum_{i=1}^n A_i \beta_i
\text{e}^{-\beta_i x}, \;x>0,$ with $A_i>0, \beta_i>0$ for
$i=1,2\cdots,n$, and $\sum_{i=1}^n A_i=1$.

 $\bullet$\;Gamma distribution with density:
$f(x)=\frac{x^{c-1}e^{-x/\beta}}{\Gamma (c)\beta^c}, \;x>0,$ with
$\beta>0$, $0<c\le 1$.

The following are several important examples of spectrally negative
L\'evy processes with completely monotone densities and that satisfy
$\int_0^{\infty}\pi(x)dx=\infty$ (cf. [19, 29]):

$\bullet$\; $\alpha$-stable process with L\'evy density:
$\pi(x)=\lambda x^{-1-\alpha},\; x>0$ with $\lambda>0$ and
$\alpha\in (0,1)\cup (1,2);$

$\bullet$\; One-sided tempered stale process (particular cases
include gamma process ($\alpha=0$) and inverse Gaussian process
($\alpha=\frac12 $)) with L\'evy density: $\pi(x)=\lambda
x^{-1-\alpha}e^{-\beta x},\; x>0$ with $\beta, \lambda>0$ and $-1\le
\alpha <2;$

$\bullet$\; The associated parent process with L\'evy density:
$\pi(x)=\lambda_1 x^{-1-\alpha}e^{-\beta x}+\lambda_2
x^{-2-\alpha}e^{-\beta x},$ $ x>0$ with $\lambda_1, \lambda_2>0$ and
$-1\le \alpha <1.$

More examples can be found in recent paper of Jeannin and Pistorius
[22, Example 2.4].

The following result can be found in Loeffen and Renaud [30]:

 \begin{lemma}\label{le5-1} Suppose the tail of the L\'evy measure is
log-convex, then, for all $\delta\ge 0$, $W^{(\delta)}$ has a
log-convex first derivative.
\end{lemma}

 \begin{theorem}\label{thrm5-1}  Suppose the tail of the L\'evy measure is
log-convex, then, for all $\delta\ge 0$, $V(x,b^*)$ is a concave
function on $(0, b^*)$, where $b^*$ is the solution of
$V'(b^*,b^*)=1$.
 \end{theorem}
{\bf Proof} \ \ Differentiate  (\ref{Thre-eq1}) with respect to $x$,
and then set $b=b^*$ yields
\begin{equation}
 V'(x,b^*)=\frac{W^{(\delta)'}(x)}{{W^{(\delta)}}'(b^*)},\; 0\le
x \le b^*.  \label{opt-eq1}
\end{equation}
 Since $b^*$ is the value of $b$ that
maximizes $V(x,b)$, i.e. $b^*$ is the value where
${W^{(\delta)}}'(b)$ attains its global minimum, and thus
${W^{(\delta)}}'(x)$ is decreasing on $(0,b^*)$. It follows that
$V'(x,b^*)$ is decreasing on $(0,b^*)$, which implies $V(x,b^*)$ is
a concave function on $(0, b^*)$.

\begin{theorem}\label{thrm5-2}  Suppose that the  L\'evy density $\pi$ is a
completely monotone function on $(0,\infty)$ and that $\delta>0$.
Then $V(x,b^*)$ is a concave function on $(b^*, \infty)$, where
$b^*$ is the solution of $V'(b^*,b^*)=1$.\
\end{theorem}
 {\bf Proof}\ \
We  prove the theorem in three cases.

Case 1: $\Pi(0,\infty)<\infty$ and $\sigma=0$. Using
(\ref{Thre-eq5}) and repeating    the proof of Lemma 8 in Kyprianou,
Loeffen and P\'erez [28] we get $V''(x,b^*)\le 0, x>b^*.$

Case 2:  $\Pi(0,\infty)=\infty$ and $\sigma=0$.   By Bernstein's
theorem we can express the function $\pi$ in the form:
$$\pi(x)=\int_0^{\infty}e^{-ux}\mu(du),$$
where $\mu$ is a measure on $(0,\infty)$. It is well known that
$\pi$ can be approximated arbitrarily closely by a hyper-exponential
density (by approximating the measure $\mu$ by sums of point
masses). It follows from Jeannin and Pistorius [22] that $X$ can be
approximated by  a sequence of approximating processes
$(X^{(n)})_{n\ge 1}$. The approximating density equal to
$$\pi_n (x)=\sum_i e^{-x u_i}\Delta_i, $$ where
$(u_i)_{i}=(u_i^{(n)})_{i}$ is the finite partitions of
$(0,\infty)$, and  $(\Delta_i)_{i}=(\Delta_i^{(n)})_{i}$ is finite
sets of positive weights. For a given $n$, the partitions
$(u_i)_{i}$  and the weights  $(\Delta_i)_{i}$
 are satisfy certain conditions. The
approximating process $X^{(n)}$ constructed in this way can be shown
to converge weakly to X. For more details, see Jeannin and Pistorius
[22, $\S$3.2]. For each $n\ge 1$, we find that
$\int_0^{\infty}\pi_n(x)dx<\infty$ and $\pi_n$ is complete monotone
on $(0,\infty)$. Let $V_n$ be the  expected discounted value of
dividend payments corresponding to the L\'evy process
$(X^{(n)})_{n\ge 1}$.  From the Case 1, we know that  $V_n(x,b_n^*)$
is a concave function on $(b_n^*, \infty)$, where $b_n^*$ is the
solution of $V_n'(b_n^*,b_n^*)=1$. The result follows, since
$\lim_{n\to\infty}V_n=V$, $\lim_{n\to\infty}V'_n=V'$ and the limit
of a pointwise convergent sequence of concave functions is concave.

Case 3:  $\sigma\neq0$.  We consider the following approximating
process $(X^{(n)})_{n\ge 1}$ (cf. Loeffen and Renaud [30]): The
L\'evy triplet is $(a_n,0,\Pi_n)$, where $a_n=a+\frac12\sigma^2 n
e^{-n}(n+1)$ and, $\Pi_n$ is defined by
$$\Pi_n(x,\infty)=\Pi(x,\infty)+\frac12\sigma^2n^2e^{-nx}.$$
For all $\theta\ge 0$,
$\lim_{n\to\infty}\psi_n(\theta)=\psi(\theta)$, and thus, by the
continuity theorem for Laplace transforms,
$\lim_{n\to\infty}W_n^{(\delta)}(x)=W^{(\delta)}(x)$ for all $x\ge
0$. Let $V_{n}$ be the expected discounted value of dividend
payments corresponding to the L\'evy process $X^{(n)}$. It is easy
to see that $\Pi_n$ has a completely monotone density and thus from
the Case 1 and Case 2, we know that $V_{n}(x,b_{n}^*)$ is a concave
function on $(b_{n}^*, \infty)$, where $b_{n}^*$ is the solution of
$V_{n}'(b_{n}^*,b_{n}^*)=1$. Because $\lim_{n\to \infty}V_{n}=V$,
$\lim_{n\to \infty}V'_{n}=V'$, the result follows. This completes
the proof of Theorem 5.2.

Combining Theorems 5.1 and 5.2 we get the main result of this paper
which established by Kyprianou, Loeffen and P\'erez [28] by using an
alternative argument:

 \begin{theorem}\label{thrm5-3} Suppose that the  L\'evy measure $\Pi$ has a
completely monotone density and that $\delta>0$. Then $V(x,b^*)$ is
a concave function on $(0, \infty)$, where $b^*$ is the solution of
$V'(b^*,b^*)=1$. Consequently, the threshold strategy  with
threshold $b^*$ is the optimal dividend strategy.
\end{theorem}

\noindent {\bf Remark 5.1.} \   If there were no restrictions on the
admissible dividend rate  $r(t)$, or, $\alpha =c$ in the case of $X$
has paths of bounded variation, then the threshold becomes a
barrier. Suppose that the L\'evy measure $\Pi$ has a completely
monotone density, then barrier strategy at $b^*$
is an optimal strategy. This result can be found in Loeffen [29].  Closely related works
in the literature are
Kyprianou et al. [26] and Yin and Wang [35].

\noindent {\bf Remark 5.2.} \ We can conclude that  the threshold
strategy is the optimal dividend strategy for the compound Poisson
risk model or the compound Poisson risk model perturbed by Brownian
motion where the claims have a distribution with a completely
monotone probability density function, which extended the
result in Gerber and Shiu [17] and Yin and Yuen [36].

\vskip0.3cm

\noindent{\bf Acknowledgements}\; The authors would like to thank the referee for many helpful suggestions and comments. The research of Chuancun Yin  was supported by
the National Natural Science Foundation of China (No.10771119, No.11171179) and
the Research Fund for the Doctoral Program of Higher Education of
China (No.20093705110002). The research of Kam C. Yuen was supported by a university research
 grant of the University of Hong Kong.

\end{document}